\documentclass{article}
\usepackage{graphicx}
\usepackage{epsfig}
\usepackage{amsfonts}
 \usepackage{amssymb}
\usepackage[dvips]{color}

\usepackage[utf8]{inputenc}
\usepackage[english]{babel}
\usepackage{amsmath,amsfonts,amssymb,amsthm}
\usepackage[left=2cm,right=2cm,top=2cm,bottom=2cm]{geometry}
\usepackage{epsfig}
\date{\today}

\usepackage{amsmath}
\usepackage{amssymb}
\usepackage[hang,nooneline,scriptsize]{subfigure}

\date{\today}

\newcommand{\insertplot}[5]{\begin{figure}
 \hfill\hbox to 0.05in{\vbox to #5in{\vfill
 \inputplot{#1}{#4}{#5}}\hfill}
 \hfill\vspace{-.1in}
 \caption{#2}\label{#3}
 \end{figure}}
 \newcommand{\inputplot}[3]{
 \special{ps: plotfile #1}
}
\newcounter{fig}

\newcommand{\ee}{\end{equation}}
\newcommand{\eea}{\end{eqnarray}}
\newcommand{\be}{\begin{equation}}
\newcommand{\bea}{\begin{eqnarray}}

\begin{document}

 \title{Spontaneous scalarization of self-gravitating magnetic fields} 

\author{
{\large Yves Brihaye}$^{(1)}$, {\large Rog\' erio Capobianco}$^{(2)}$  and 
{\large Betti Hartmann}$^{(2),(3),(4), (5)}$
\\ 
\\
$^{\rm (1)}$\normalsize{Service de Physique de l'Univers, Champs et Gravitation, Universit\'e de
Mons, Mons, Belgium}\\
{$^{\rm (2)}$\normalsize{Instituto de F\'isica de S\~ao Carlos, Universidade de S\~ao Paulo, S\~ao Carlos, S\~ao Paulo 13560-970, Brazil}}\\
{$^{\rm (3)}$\normalsize{Institut f\" ur Physik, Carl-von-Ossietzky Universit\"at Oldenburg, 26111 Oldenburg, Germany}}\\
{$^{\rm (4)}$\normalsize{Department of Physics and Earth Sciences, Jacobs University Bremen, 28759 Bremen, Germany 
}}\\
{$^{\rm (5)}$\normalsize{Department of Mathematics, University College London, Gower Street, London, WC1E 6BT, UK
}}}

\maketitle 
\begin{abstract} 
In this paper, we study the spontaneous scalarization of an extended, self-gravitating system which is static, cylindrically symmetric and possesses electromagnetic fields.  
We demonstrate that a real massive scalar field condenses on this Melvin magnetic universe
solution when introducing a non-minimal coupling between the scalar field and (a) the magnetic field and (b) the curvature of the space-time, respectively. We find that in both cases, the solutions exist on a finite interval of the coupling constant and that solutions with a number of nodes $k$ in the scalar field exist. For case (a) we observe that 
the intervals of existence are mutually exclusive for different $k$.   

\end{abstract}

\section{Introduction}
Multi-Messenger observations of compact objects allow to test General Relativity (GR) and its extensions to high precision
now and in the future. As such, re-newed interest in testing No-hair and/or uniqueness theorems for  black holes has appeared. While work in the 1990s has mainly been devoted to the construction of  ``hairy'' black holes in the context of General Relativity suplemented with non-linear matter fields appearing in particle physics models, recent activity has focused on the extention of the gravity part of the model, e.g.  by adding a non-minimal interaction between 
higher order curvature terms  and extra gravitational fields see e.g. \cite{Herdeiro:2015waa, Sotiriou:2015pka} for reviews.

In numerous of these extended gravity models new black hole solutions with non-trivial fields 
on the horizon that vanish asymptotically have been shown to exist. 
In fact, these new black hole solutions appear for specific intervals of the non-minimal coupling. Outside of this interval, the black hole solutions are equivalent to 
the standard black hole solutions that fulfill the No-hair theorems, i.e. are equivalent to either the Schwarschild, Reissner-Nordstr\"om or Kerr (-Newman) solution. 
In these extended models black holes are hence said to ``scalarize spontaneously'' \cite{Doneva:2017bvd, Silva:2017uqg, Antoniou:2017acq} in the case of  non-minimal coupling to a scalar field or ``vectorize spontaneously'' \cite{Ramazanoglu:2017xbl,Ramazanoglu:2018tig,Ramazanoglu:2019gbz, Barton:2021wfj} in the case of non-minimal coupling to abelian gauge fields. 

However, the idea of spontaneous scalarisation is not specific to black holes, but has been shown to appear also for other compact objects such as boson stars \cite{Whinnett:1999sc,    
Alcubierre:2010ea, Brihaye:2019puo} and neutron stars \cite{Damour:1993hw}. 

In this paper, we want to show that spontaneous scalarization does not need a compact object, but can also appear in extended self-gravitating systems. For that we study the Melvin solution that describes an extended electromagnetic field  kept together by its own gravity \cite{melvin} and was shown to be stable in the context of GR \cite{Thorne:1965}. Charged black holes embedded in such a magnetic universe
have been study recently in \cite{Santos:2021nbf} and it has been shown that they can carry minimally coupled,
complex and ungauged scalar hair.

The electromagnetic field of the ``pure‘‘ Melvin solution points into the direction of the symmetry axis and
the solution is essentially characterized by the absolute value of this field on the symmetry axis.
As such it is a cylindrically symmetric gravitating system (for a review see e.g. \cite{Bronnikov:2019clf}). 
The most studied cylindrically symmetric extended self-gravitating system is surely the cosmic string, a topological defects that might have formed in the primordial
universe \cite{Hindmarsh:1994re}.  However, magnetic fields in elongated regions were observed astrophysically in so-called radio relics.
Radio relics are diffuse radio sources in galaxy clusters. These sources are not associated to any cluster galaxy  \cite{kierdorf2017relics} and have been categorized into three groups: radio gischt, radio phoenix and active galactic nucleus (AGN) relics \cite{2004rcfg.proc..335K}. The radio gischt are mostly found in the outskirts of galaxies and are elongated arc-like radio sources with sizes of up to 2 Mpc. Observations give support to the hypothesis that they trace shock fronts in which particles are accelerated via the diffuse shock accleration mechanism. One interesting case of gischt-like sources are so-called ``double-relics‘‘. In this case two relics are diametrically located on both sides of the cluster center, see e.g. \cite{van2011using} and references therein.    
Due to the improvement in instrument sensitivity the number of detections of radio relics has grown dramatically in the last decade. Large cosmological simulations that include radio emissions from shocks suggest that these structures should form frequently, see e.g. \cite{nuza2011radio,van2011using} and references therein.

Motivated by the existence of elongated magnetic fields in the universe and the aim to extend the process
of spontaneous scalarization to non-compact self-gravitating objects, we study the Einstein-Maxwell
model and add a massive, real scalar that is non-minimally coupled to the system. In order to understand
the effects of the non-minimal coupling we study two different scenarios separately: (a) the non-minimal coupling to the electromagnetic field and (b) the non-minimal coupling to the Gauss-Bonnet curvature term.
These two coupling options have been used extensively in the recent construction of black holes with scalar hair. 

Our paper is organized as follows: in Section 2, we give the model and Ansatz and also discuss the small scalar field limit. In Section 3 we present our results for the case of scalar-magnetic field coupling, while
Section 4 is concerned with the scalar-gravity case. 
We conclude in Section 5.

\section{The model and Ansatz}
In this paper, we study a scalar-tensor gravity model with the following action
\begin{equation}
\label{action}
S=\int {\rm d}^4x  \ \sqrt{-g} \left[\frac{\cal R}{16\pi G}  +  \phi^2\left(\alpha F_{\mu\nu}  F^{\mu\nu} + \gamma {\cal G}\right)  +  \frac{1}{2}\partial_{\mu} \phi \partial^{\mu} \phi  - \frac{m^2}{2}\phi^2 - \frac{1}{4} F_{\mu\nu} F^{\mu\nu} \right]  \ ,
\end{equation}
where ${\cal R}$ is the Ricci scalar, ${\cal G}$ the Gauss-Bonnet term, $F_{\mu\nu}=\partial_{\mu} A_{\nu}  -  \partial_{\nu} A_{\mu}$ the field strength tensor
of a U(1) gauge field $A_{\mu}$
and $\phi$ a real-valued scalar field with mass $m$ that
is coupled to the Maxwell invariant  $F_{\mu\nu} F^{\mu\nu}$ as well as the Gauss-Bonnet term ${\cal G}$ given by
\begin{equation}
{\cal G} = ({\cal R}^{\mu \nu \rho \sigma} {\cal R}_{\mu \nu \rho \sigma} - 4 {\cal R}^{\mu \nu} {\cal R}_{\mu \nu} + {\cal R}^2 ) 
\end{equation}
via the couplings $\alpha$ and $\gamma$, respectively.  The equations of motion then read
\begin{equation}
\label{eq:eq1}
\square\phi+\left(2\alpha F_{\mu\nu}  F^{\mu\nu} +  2\gamma {\cal G}-m^2\right) \phi =0   \ \ , 
\end{equation}
and 
\begin{equation}
\label{eq:eq2}
\partial_{\mu} \left(\sqrt{-g} (1-4\alpha\phi^2) F^{\mu\nu}\right) = 0  \ \ , \ \ 
G_{\mu\nu} = -8\pi G \left(T^{(A)}_{\mu\nu} +   T^{(\phi)}_{\mu\nu}\right) \ \ , \
\end{equation}
where the energy-momentum tensor components of the gauge field and scalar field read, respectively:
\begin{equation}
T^{(A)}_{\mu\nu} =  \left(\frac{1}{4}-\alpha \phi^2\right)  \left( F_{\mu\sigma}  F_{\nu}^{\sigma} - \frac{1}{4} g_{\mu\nu} F_{\alpha\beta}
F^{\alpha\beta}  \right)  \ \ , \ \     
\end{equation}
\begin{equation}
T^{(\phi)}_{\mu\nu}=  \partial_{\mu} \phi \partial_{\nu} \phi - g_{\mu\nu}\left(\frac{1}{2} \partial_{\sigma} \phi \partial^{\sigma} \phi  + \frac{m^2}{2} \phi^2 \right)   - \gamma\left(g_{\mu\sigma} g_{\nu\lambda} + g_{\nu\sigma} g_{\mu\lambda}\right) \eta^{\sigma\alpha\gamma\delta} \eta^{\iota\lambda\kappa\rho} {\cal R}_{\gamma\delta\kappa\rho} D_{\alpha} D_{\iota} (\phi^2)   \ . 
\end{equation}

In this paper, we would like to discuss the scalarization of self-gravitating solutions of the Einstein-Maxwell equations. We assume staticity and
cylindrical symmetry and hence choose the following Ansatz for the metric, gauge and scalar field~:
\begin{equation}
 {\rm d}s^2 = N^2  {\rm d}t^2 - H^2  {\rm d}\rho^2 - L^2  {\rm d}\varphi^2 - K^2  {\rm d}z^2 \ \ , \ \ A_{\mu}  {\rm d}x^{\mu} = A(\rho) {\rm d}\varphi \ \ , \ \
  \phi= \phi(\rho)
\end{equation}
where the metric functions $N$, $H$, $J$, $K$ depend only on $\rho$. In the following, we will now fix the gauge by imposing $H(\rho)=1$, which implies $K(\rho)\equiv N(\rho)$. Inserting the Ansatz into the equations of motion (\ref{eq:eq1}), (\ref{eq:eq2}) we note that the Maxwell equation can be integrated separately, leading to~:
\be
\label{eq:maxwell}
             A' = B_0\frac{L}{(1-4 \alpha \phi^2)N^2}  \ , 
\ee
where $B_0$ is an integration constant. The magnetic field of the solution, which points in the direction
of the $z$-axis, is then given by~:
\begin{equation}
\label{eq:magnetic_field}
{\cal B}=-\frac{A'}{L} = - \frac{B_0}{(1-4\alpha \phi^2) N^2}  \ . 
\end{equation} 
The remaining equations read~:
\begin{eqnarray}
\label{eq1}
 \frac{N''}{N} + \frac{(N')^2}{2 N^2} &=& \frac{\kappa}{4}    (-\epsilon_s + \epsilon_v - U )  + 4\kappa\gamma     {\cal F}_1 \         ,  \\
 \label{eq2}
	\frac{L''}{L}		+ \frac{L'N'}{LN}	- \frac{(N')^2}{2 N^2} &=&  \frac{\kappa}{4}  (-\epsilon_s -3 \epsilon_v - U ) - 4\kappa \gamma {\cal F}_2 \ , \\
	\label{eq3}
	\phi'' + \left(\frac{L'}{L} + \frac{2 N'}{N} \right)\phi' &=& \phi\left(m^2 - 4 \alpha \frac{(A')^2}{L^2} - 16\gamma \frac{L'' N'^2 + 2 L' N' N''}{LN^2} \right)   \ , 
\end{eqnarray}
where we have used the following abbreviations
\begin{equation}
     \kappa = 16 \pi G \ \ , \ \      \epsilon_s =  \frac{(\phi')^2}{2} \ \ , \ \ \epsilon_v = \frac{(A')^2}{2 L^2} (1-4 \alpha \phi^2) \ \ , \ \ U = \frac{m^2}{2} \phi^2  \ , 
\end{equation}
as well as 
\begin{eqnarray}
 {\cal F}_1 &=& \frac{N'^2}{N^2} (\phi \phi'' + \phi'^2) + \frac{2 N'N''}{N^2} \phi \phi'  \ , \nonumber \\
 {\cal F}_2 &=& \left(\phi \phi''  +  \phi'^2 \right)   \left( \frac{N'^2}{N^2} - 2 \frac{L' N'}{LN}  \right)  + \phi \phi' \left(\frac{2N' N''}{N^2} -
2 \frac{L' N''}{LN} - 2 \frac{L'' N'}{LN}\right) 
\end{eqnarray}
and the prime denotes derivative with respect to $\rho$. Moreover, we have a constraint, which reads
\begin{equation}
\label{eq:bianchi}
\frac{N'^2}{2N^2} + \frac{N' L'}{NL} = \frac{\kappa}{4} \left(  \epsilon_s +  \epsilon_v - U   \right) + \kappa\gamma {\cal F}_3  \ , \   {\cal F}_3=12\gamma \kappa \phi\phi' \left( \frac{L' N'^2}{L N^2}\right)   \ .
\end{equation}

The system has to be solved for $\rho \in [0,\infty[$ with the following boundary conditions which guarantee the regularity at origin and the localization of the solution~:
\be
\label{eq:bc}
    N(0)=1 \ \ , \ \ N'(0) = 0 \ \ , \ \ L(0) = 0 \ \ , \ \ L'(0) = 1 \ \ , \ \ \phi'(0) = 0 \ \ , \ \ \phi(\rho \to \infty) = 0 \ \ .
\ee
Note that $\kappa$ and $m$ can be set to unity by appropriate rescalings of the fields and of the radial variable, respectively. 

In the vacuum case, i.e. for $\phi\equiv 0$, $B_0=0$, the equations of motion have well known solutions first given in \cite{levi_civita}. For the boost-symmetric case,  these are~:
\begin{itemize}
\item $N\sim 1$, $L\sim \beta r$: this is a locally flat space-time which globally possesses a deficit angle $\Delta=2\pi (1-\beta)$. The metric
describes e.g. the (asymptotic) space-time of a cosmic string (see e.g. \cite{Hindmarsh:1994re} and references therein). 
\item $N\sim r^{2/3}$, $L\sim r^{-1/3}$: this space-time obviously does not fulfill the regularity conditions (see (\ref{eq:bc})) on the axis, however, is important in the following in the description of  the space-time away from the sources of the gravitational
field.
\end{itemize}

\subsection{Small and vanishing scalar field}
In the case of vanishing scalar field, i.e. for the case $\phi\equiv 0$, a combination of the equations (\ref{eq1}), (\ref{eq2}) and (\ref{eq:bianchi}) shows that the metric functions have to fulfill $N' \propto L$.
This clearly excludes the string-type solution far away from the magnetic field, while the vacuum solution
with $L\sim \rho^{-1/3}$, $N\sim \rho^{2/3}$ fulfills this requirement. 
In fact, the solution can be given in closed form and is often referred to as the magnetic Melvin universe \cite{melvin}. In order to discuss the scalar field in this background, it is convenient to adopt Weyl-type coordinates
with ${\rm d} \rho = N {\rm d} r$. The solution then reads~:
\begin{equation}
{\rm d}s^2 = N^2 \left(  {\rm d}t^2 -   {\rm d}r^2 - d\varphi^2\right) - \frac{r^2}{N^2}  {\rm d}\varphi^2  \ \ , \ \    F_{r\varphi}=\frac{B_0 r}{N^2}  \ , \ \ N=\left(1+\frac{1}{4} B_0^2 r^2\right) \ . 
\end{equation}
The relation between $\rho$ and $r$ is
\begin{equation}
\rho= r + \frac{1}{12} B_0^2 r^3 \ \ , \ \ {\textbf{resp.}} \ , \    r=\left(\frac{6\rho}{B_0^2} (1+ \sqrt{\Sigma})\right)^{1/3} + \left(\frac{6\rho}{B_0^2} (1- \sqrt{\Sigma})\right)^{1/3}  \ , \ \Sigma=1 +\frac{1}{27 B_0^2 \rho^2}  \ .
\end{equation}

In the following, we will use the coordinate $r$ to study the scalar field
equation in the background of this solution. The scalar field equation (see (\ref{eq:eq1})) then reads
\begin{equation}
\label{eq:scalar_probe}
\frac{1}{r}\partial_r \left(r\partial_r \phi\right) +  (2\alpha F_{\mu\nu}  F^{\mu\nu} + 2\gamma {\cal G}-m^2)N^2 \phi =0  \ \ \ \ , \ \ \  {\cal G}  = \frac{(3 B_0^4 r^4 - 24 B_0^2 r^2 + 16) B_0^4}{4N^8}  \ , \   F_{\mu\nu}  F^{\mu\nu}  = \frac{2B_0^2}{N^4}  \ .
\end{equation}
The general solution to this equation can only be found numerically, but we can understand the behaviour of the solutions when looking at the asymptotic behaviour of the scalar field.  We will discuss the two cases $\gamma=0$ and $\alpha=0$ separately now. 
\begin{enumerate}
\item {\bf $\gamma=0$} \\
For $r\ll 1$, we can approximate $N^{\pm 2}\approx 1\pm B_0^2 r^2/2$ and the equation (\ref{eq:scalar_probe}) becomes
\begin{equation}
\label{eq:scalar_probe_rsmall}
\frac{1}{r}\partial_r \left(r\partial_r \phi\right)  - {\cal A}_0 \phi - {\cal A}_2 r^2 \phi = 0 \ \ , \ \  {\cal A}_0 = m^2 - 4 \alpha B_0^2 \ \ , \ \ 
{\cal A}_2=\frac{m^2 B_0^2}{2} + 2 \alpha B_0^4  \ .
\end{equation}
Introducing $z=\sqrt{{\cal A}_2} r^2$ and defining $\phi=\exp(-z/2)\chi$, we obtain
\begin{equation}
\label{eq:scalar_probe_rsmall_z}
z\ddot{\chi} + (1-z)  \dot{\chi} + {\cal A}\chi = 0 \ \ , \ \   {\cal A}=-\left( \frac{{\cal A}_0}{4  \sqrt{{\cal A}_2}} + \frac{1}{2}  \right) \ ,
\end{equation}
where the dot denotes the derivative with respect to $z$. This is the confluent hypergeometric equation that has as suitable solutions the
Laguerre polynomials $\chi(z)\sim {\cal L}_{{\cal A}}$. Hence for small $r$ the equation (\ref{eq:scalar_probe_rsmall}) has the solution
\begin{equation}
\label{eq:laguerre}
\phi(r\ll 1) =\phi_0 \exp\left(-\frac{\sqrt{{\cal A}_2}}{2} r^2\right) {\cal L}_{{\cal A}}\left(\sqrt{{\cal A}_2} r^2\right) \ . 
\end{equation}
For ${\cal A}\in \mathbb{N}$ the ${\cal L}_{{\cal A}}$ possess a number of nodes. This suggests that we should also be able to construct scalar field solutions that possess a number $k$ of nodes, a conclusion that we have confirmed by an explicit numerical construction, see below. In fact, using these arguments, we can give a rough approximation
of the critical value of $\alpha$ to obtain solutions. From the requirement that ${\cal A}=k$, $k=0,1,2,...$, we find that 
\begin{equation}
\label{eq:alpha_cr}
\alpha \gtrsim \frac{m^2}{4 B_0^2} \ \ {\text{for}} \ \  k=0,1,2,3,... \\  . 
\end{equation}

For $r\gg 1$ we introduce $y=r^3$  and the equation (\ref{eq:scalar_probe}) becomes a modified Bessel equation of the form
\begin{equation}
y^2 \frac{{\rm d}^2 \phi}{{\rm d} y^2} + y \frac{{\rm d} \phi}{{\rm d} y}  - \frac{m^2 B_0^4}{144} y^2 \phi = 0 \ 
\end{equation}
such that the asymptotic decay of the solution is
\begin{equation}
\phi(r\gg 1)\sim K_0 \left(\frac{m B_0^2}{12} r^3 \right) \sim  r^{-3/2} \exp(-r^3)   \ .
\end{equation}  
This analysis also clearly demonstrates why it is necessary to have a mass term for the scalar field.  For $m=0$, as is well known, the
scalar field would behave like $\phi(r)\sim \ln(r)$ asymptotically and would hence not be localized.

\item {\bf $\alpha=0$} \\
In this case equation (\ref{eq:scalar_probe}) becomes~:
\begin{equation}
\label{eq:scalar_probe_rsmall}
\frac{1}{r}\partial_r \left(r\partial_r \phi\right)  - {\cal C}_0 \phi - {\cal C}_2 r^2 \phi = 0 \ \ , \ \  {\cal C}_0 = m^2 - 8 \gamma B_0^4 \ \ , \ \ 
{\cal C}_2=\frac{m^2 B_0^2}{2} + 24 \gamma B_0^6  \ .
\end{equation}
With similar substitutions as above, we find 
\begin{equation}
\phi(r\ll 1) =\phi_0 \exp\left(-\frac{\sqrt{{\cal C}_2}}{2} r^2\right) {\cal L}_{{\cal C}}\left(\sqrt{{\cal C}_2} r^2\right) \ , \  {\cal C}=-\left( \frac{{\cal C}_0}{4  \sqrt{{\cal C}_2}} + \frac{1}{2}  \right) \ .
\end{equation}
Again, the analysis suggests that radially excited solutions should be present and we can give a rough approximation
of the critical value of $\gamma$ to obtain solutions. From the requirement that ${\cal C}=k$, $k=0,1,2,...$, we find that 
\begin{equation}
\label{eq:gamma_cr}
\gamma \gtrsim \frac{m^2}{8 B_0^4} \ \ {\text{for}} \ \  k=0,1,2,3,... \\  . 
\end{equation}

For $r\gg 1$ the behaviour is exactly as in the $\gamma=0$ because it is the mass term that determines the asymptotic regime in both cases.

\end{enumerate}

\section{Scalar-magnetic field coupling}
\begin{figure}[h!]
\begin{center}
{\includegraphics[width=10cm]{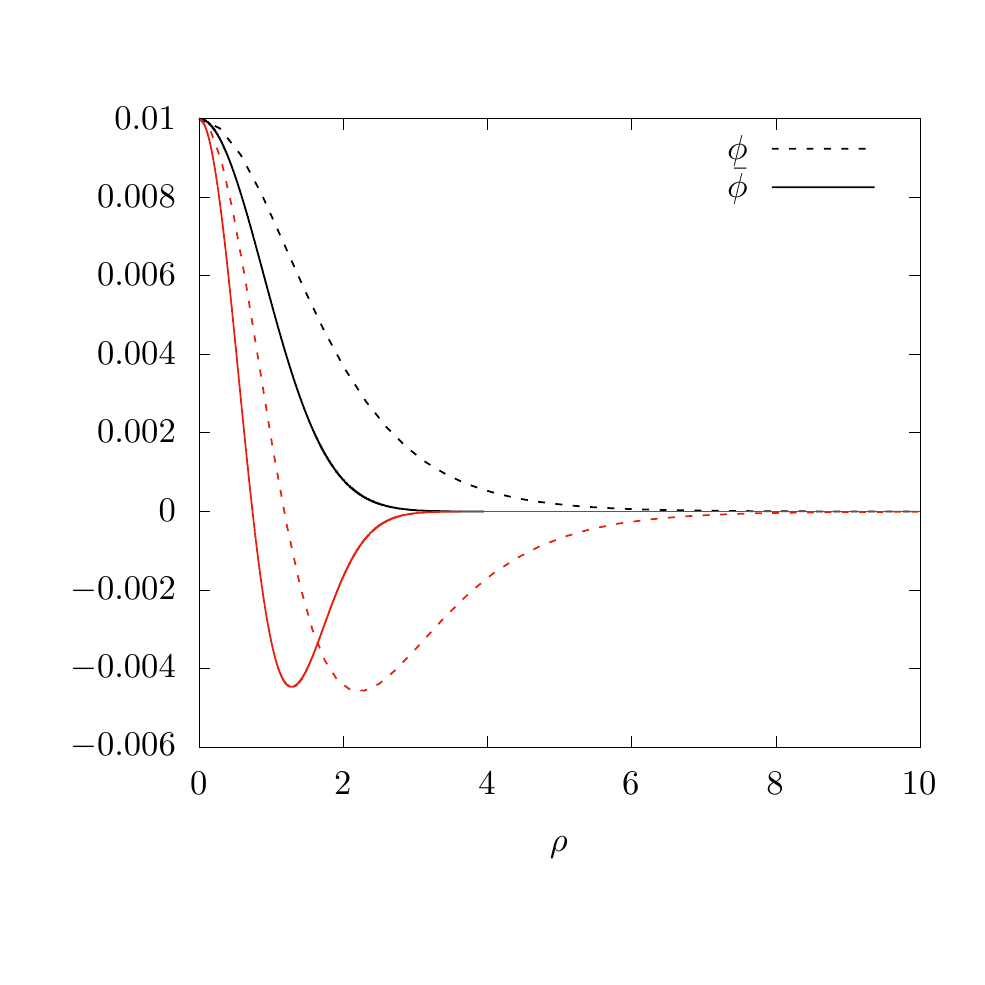}}
\vspace{-1cm}
\caption{We compare the profiles of the analytical approximation $\bar{\phi}$ (see (\ref{eq:laguerre})) (solid) for the scalar field 
with the full numerical solution $\phi$ (dashed) for $k=0$ (black) and $k=1$ (red)  nodes for $B_0=1.0$ and $\phi(0)=0.01$.  These choices correspond to $\alpha=0.55$ for $k=0$ and $\alpha=1.40$ in the case $k=1$, respectively. Note that we have used $\rho\approx r$ for this plot, which is a good approximation for small values of the radial coordinate.
\label{fig:comparison_nodes}
}
\end{center}
\end{figure}

Here, we would like to discuss the case $\gamma=0$, i.e. we consider only the non-minimal coupling between the gauge field and the scalar field.
As stated above, we can choose appropriate scalings to set $\kappa=m\equiv 1$ without loosing generality. The parameters to be
varied in the following are then the non-minimal coupling constant $\alpha$ and the absolute value of the magnetic field strength $B_0$. 
As discussed above for small scalar fields, we expect solutions with scalar field nodes to be present in our system. 
We, indeed, have confirmed this numerically. In Fig. \ref{fig:comparison_nodes}, we compare the
analytical expression (\ref{eq:laguerre}) (denoted $\bar{\phi}$ and given in solid) with the numerical solutions
of the full set of equations for $B_0=1$ and $\phi(0)=0.01$ (dashed) for the solution with no nodes ($k=0$) and that
with one node ($k=1$). As expected, the approximation is not perfect, but gives a good idea of the qualitative behaviour of the functions.  We also find that the approximation gives a good order of magnitude approximation of the
location of the zeros of the scalar field function. We give some values for the location of the nodes of the $k=2$
solution, i.e. the solution with two nodes, in comparison to the location of the zeros of the second Laguerre polynomial
${\cal L}_2$ in Table \ref{table:comparison}. 

\begin{table}
\begin{center}
    \begin{tabular}{ | l | l | l | l | | l | l | }
    \hline \hline
    $B_0$ & $\alpha$ & $\rho_0^{(1)}$ & $\rho_0^{(2)}$ & $\bar{\rho}_0^{(1)}$ &  $\bar{\rho}_0^{(2)}$\\ \hline
    $0.05$ & $124.2$ & $4.1$ & $11.5$ & $2.6$ & $15.5$ \\ \hline 
    $0.07$ & $69.5$ & $3.8$ & $9.6$ & $2.1$ & $13.2$ \\ \hline 
    $0.2$ & $13.7$ & $2.3$ & $5.5$ & $1.2$ & $7.5$ \\ \hline 
    $0.4$ & $5.8$ & $1.5$ & $3.8$ & $0.8$ & $5.5$ \\ \hline 
    $0.5$ & $4.6$ & $1.3$ & $3.3$ & $0.6$ & $4.8$ \\ 
    \hline
    \end{tabular}
    \caption{We give the location of the zeros of the scalar field solution with $k=2$ notes, $\rho_0^{(1)}$ and $\rho_0^{(2)}$, for some values of $B_0$ and $\alpha$ and compare them with the zeros of the corresponding Laguerre polynomial
    ${\cal L}_2$, $\bar{\rho}_0^{(1)}$ and  $\bar{\rho}_0^{(2)}$, respectively. }
\end{center}
\label{table:comparison}
\end{table}

\begin{figure}[ht!]
\hspace{-1cm}
\includegraphics[width=9cm]{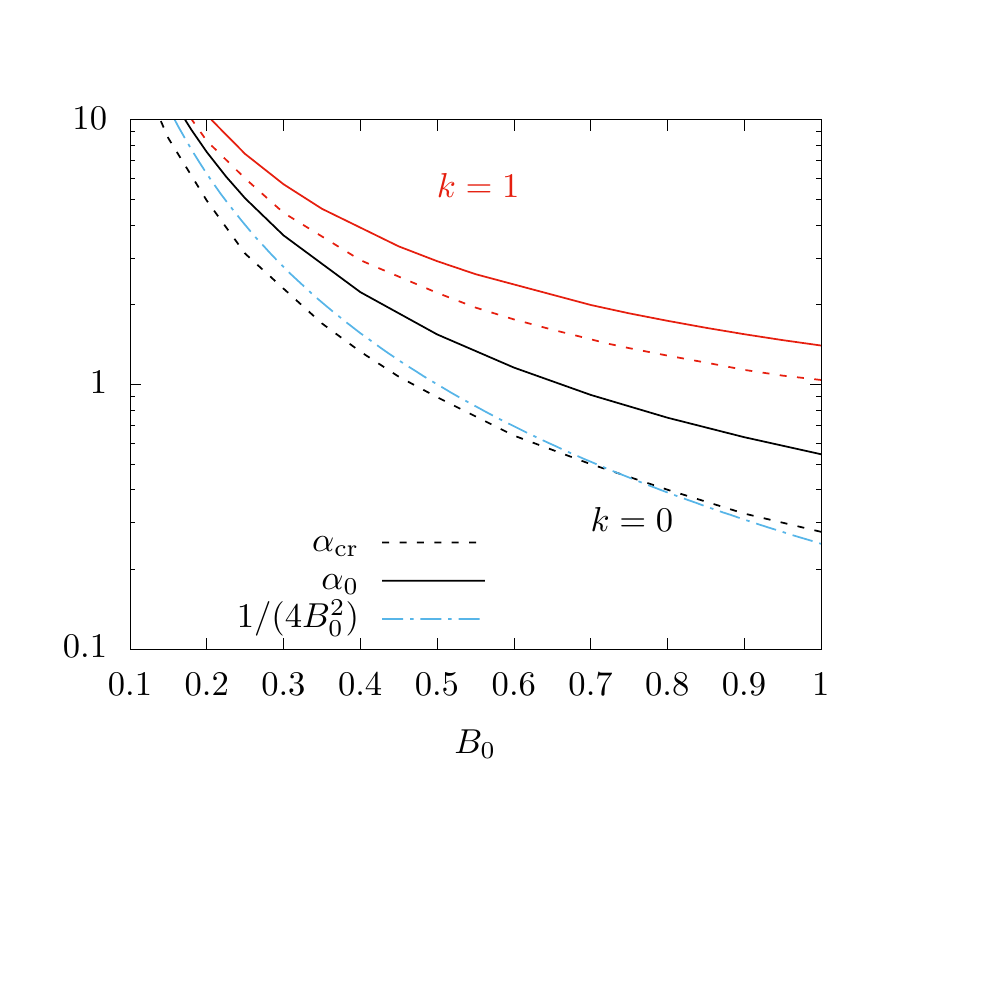}
\includegraphics[width=9cm]{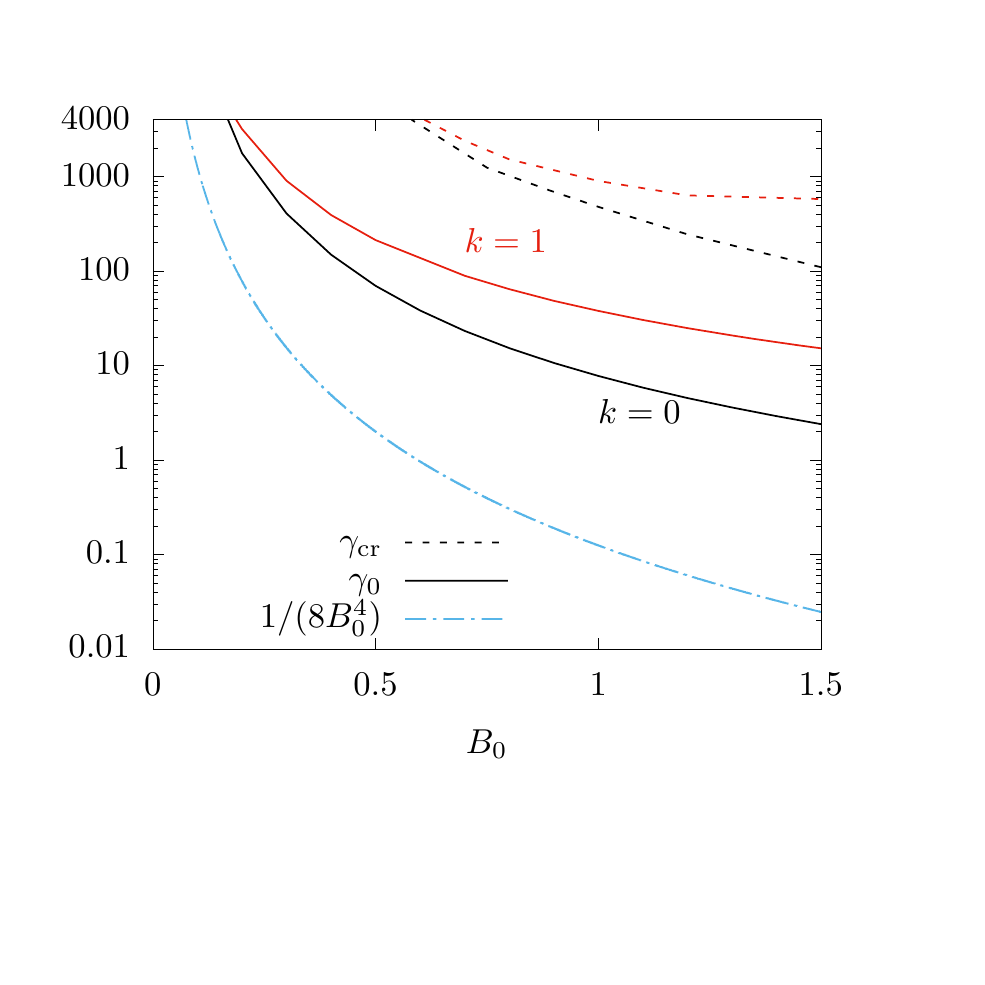}
\vspace{-2.5cm}
\begin{center}
\caption{{\it Left}: We show the values of $\alpha_0$ (solid) and $\alpha_{\rm cr}$ (dashed) between which the scalarized solutions with $k=0$ (black) and $k=1$ (red) nodes, respectively, exist in dependence on the magnetic field parameter $B_0$ for $\gamma=0$. We also give $1/(4 B_0^2)$ (dotted-dashed, blue) which is
a rough approximation of the critical value of $\alpha$ (see text for details). 
{\it Right}: We show the values of $\gamma_0$ (solid) and $\gamma_{\rm cr}$ (dashed) between which the scalarized solutions with $k=0$ (black) and $k=1$ (red) nodes, respectively, exist in dependence on the magnetic field parameter $B_0$ for $\alpha=0$. We also give $1/(8 B_0^4)$ (dotted-dashed, blue) which is
a rough approximation of the critical value of $\alpha$ (see text for details). 
\label{fig:ff2_alpha_b0}}
\end{center}
\end{figure}

As such the $k=0$ and $k=1$ in Fig.\ref{fig:ff2_alpha_b0}, respectively, refers to the solution that has no nodes in the scalar field ($k=0$) and that which has one node ($k=1)$. 
Moreover, (\ref{eq3}) suggests that a tachyonic instability appears in the system only for $\alpha > 0$. From (\ref{eq:alpha_cr}), we know that 
$\alpha \gtrsim \frac{m^2}{4 B_0}$. The curve $\alpha=\frac{m^2}{4 B_0}=\frac{1}{4 B_0}$
(remember $m\equiv 1$) is given in Fig. \ref{fig:ff2_alpha_b0} (blue, dotted-dashed) together
with the numerically determined values of $\alpha_0$ (solid) and $\alpha_{\rm cr}$ (dashed) which determine the interval
in $\alpha$ for which non-trivial scalar field solutions exist for a given value of $B_0$. Here, the value of $\alpha_{\rm cr}$
is given by the observation that there exists a value of $\alpha=1/(4\phi(0)^2)$ for which the magnetic field on the axis of symmetry (see (\ref{eq:magnetic_field})) diverges. Solutions exist for $\alpha > \alpha_{\rm cr}$. That this is closely tight to 
the fact that $\alpha$ needs to be sufficiently large to generate a tachyonic instability can be seen by noting
that the curve $\alpha=1/(4\phi(0)^2)$ is a good approximation to the $\alpha_{\rm cr}$-curve for $k=0$ and small $\alpha$. 

Increasing $\alpha$ too strongly, the scalar field can no longer be non-trivial in the space-time
and becomes identically zero due to the backreaction of the space-time. This value of $\alpha$ is  denoted by $\alpha_0$.
Note that
had we only studied the scalar field in the background of the Melvin universe, the two limiting values would be equal 
$\alpha_0=\alpha_{\rm cr}$. However, here, the backreaction of the scalar field on the space-time leads to the observation
that non-trivial scalar field solutions exist in a given interval of $\alpha$ (for a given $B_0$) rather than for
a sole value of $\alpha$. 
We observe that the interval in $\alpha$ increases with increasing magnetic field strength $B_0$. For $B_0\rightarrow 0$, our numerical results indicate that the interval shrinks to zero and both $\alpha_{\rm cr}$ as well as $\alpha_0$ tend to infinity.
This makes sense since the vanishing $B_0$ limit corresponds to Minkowski space-time (in cylindrical coordinates) and this space-time cannot be scalarized.

Comparing the scalar field solutions with a different number of nodes $k$,  we observe that for a given $B_0$, we have to choose
$\alpha$ larger to find $k=1$ solutions than $k=0$ solutions. Interestingly, the ranges of $\alpha$ for which
$k=0$ and $k=1$ solutions exist, respectively, do not overlap in the range of values of $B_0$ that we have studied here.
To state it differently~: fixing $B_0$ and $\alpha$ within the range of values given in Fig. \ref{fig:ff2_alpha_b0} we will
either obtain a scalar field solution with no nodes or one with one node, but not both at the same time.
Hence, the solutions with nodes cannot really be interpreted as the radially excited solution of the fundamental
ones, as is often done in other non-linear, self-gravitating systems. 

\section{Scalar-curvature coupling}
Here, we will set $\alpha=0$, i.e. we consider only the non-minimal coupling between the Gauss-Bonnet term and the scalar field. In this case, the equations of motion (\ref{eq:eq1})-(\ref{eq3}) have to be diagonalized with respect to the second derivatives. The explicit expressions
of these diagonalized equations are lengthy, that is why we obmit them here. However, let us state the determinant of the matrix that appears in this diagonalization~:
\begin{equation}
\Delta= 1536 \gamma^3 \kappa^2  L' N'^4\phi^3\phi' + 64\gamma^2 \kappa N'^2\phi^2\left( -\kappa L N^2\phi'^2 + L N'^2 - 4 L' N N' \right)
+ 16\gamma \kappa L N^3 N'  \phi\phi'  - L N^4  \ . 
\end{equation}

We observe that non-trivial scalar field solutions exist only for sufficiently large values of the non-minimal coupling $\gamma$, i.e. for $\gamma > \gamma_0$. The dependence of this value on $B_0$ is shown in 
Fig. \ref{fig:ff2_alpha_b0} (solid lines) for scalar field solutions with no nodes ($k=0$, black) and
one node ($k=0$, black), respectively. Again, we observe that have to choose the non-minimal
coupling large in order to obtain solutions with $k=1$ as compared to the $k=0$ case. 
Interestingly, in this case, the analytical expression for
$\gamma_0$ given by $1/(8B_0^4)$ is not as good as in the scalar-magnetic field coupling case. This is likely related to the fact that the space-time background approximation is not a good approximation in this case as the scalar field is non-minimally coupled to the curvature. 
When increasing the coupling $\gamma$, the scalar field increases in absolute value, which leads to increased backreaction of the scalar field on the space-time. In fact, we observe that there exists a maximal value of $\phi(0)$, or equivalently a maximal value of $\gamma=\gamma_{\rm cr}$ beyond which no scalarized solutions exist anymore. This is true for both the $k=0$ and the $k=1$ case. The values of $\gamma_{\rm cr}$
in dependence of $B_0$ are shown in Fig. \ref{fig:ff2_alpha_b0} (dashed lines). Note that for the
scalar-curvature case the critical value of the coupling is always larger than the value where the scalar
field vanishes identically, while for the scalar-magnetic field coupling, this is exactly opposite. 
This is related to the fact that the scalar field directly sources these fields and hence leads to an increased
repulsive effect for the magnetic fields and an increased attractive effect in the case of the curvature fields.
This also demonstrates that the two couplings are qualitatively different in nature.
Another difference to the scalar-magnetic field coupling is that the domain of existence of scalarized solutions
for $\gamma\in [\gamma_0:\gamma_{\rm cr}]$ is now not mutually exclusive for different node solutions.
The range of $\gamma$ for $k=0$ overlaps partially with the range of $\gamma$ for $k=1$, as Fig. \ref{fig:ff2_alpha_b0} clearly demonstrates, and in this overlapping region the $k=1$ solutions can be 
interpreted as the radially excited version of the $k=0$ solution. 

\begin{figure}[ht!]
\hspace{-1cm}
\includegraphics[width=9cm]{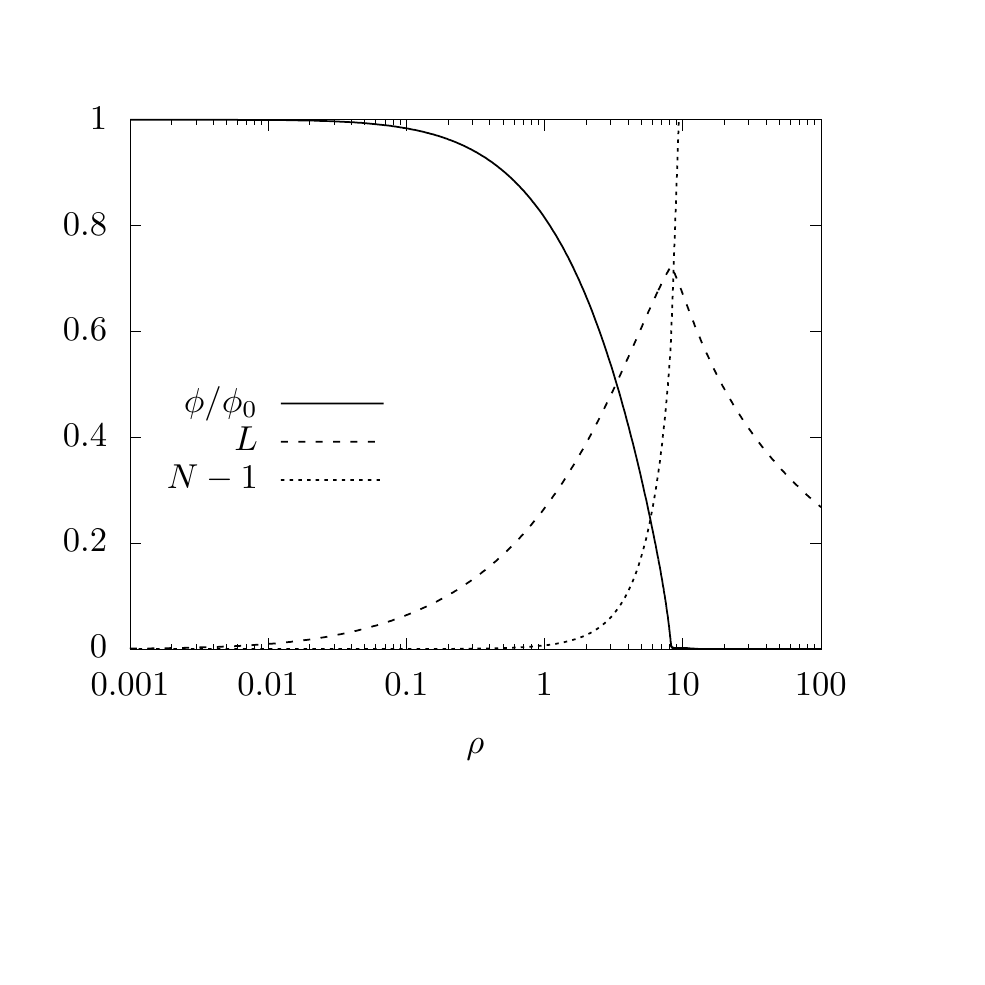}
\includegraphics[width=9cm]{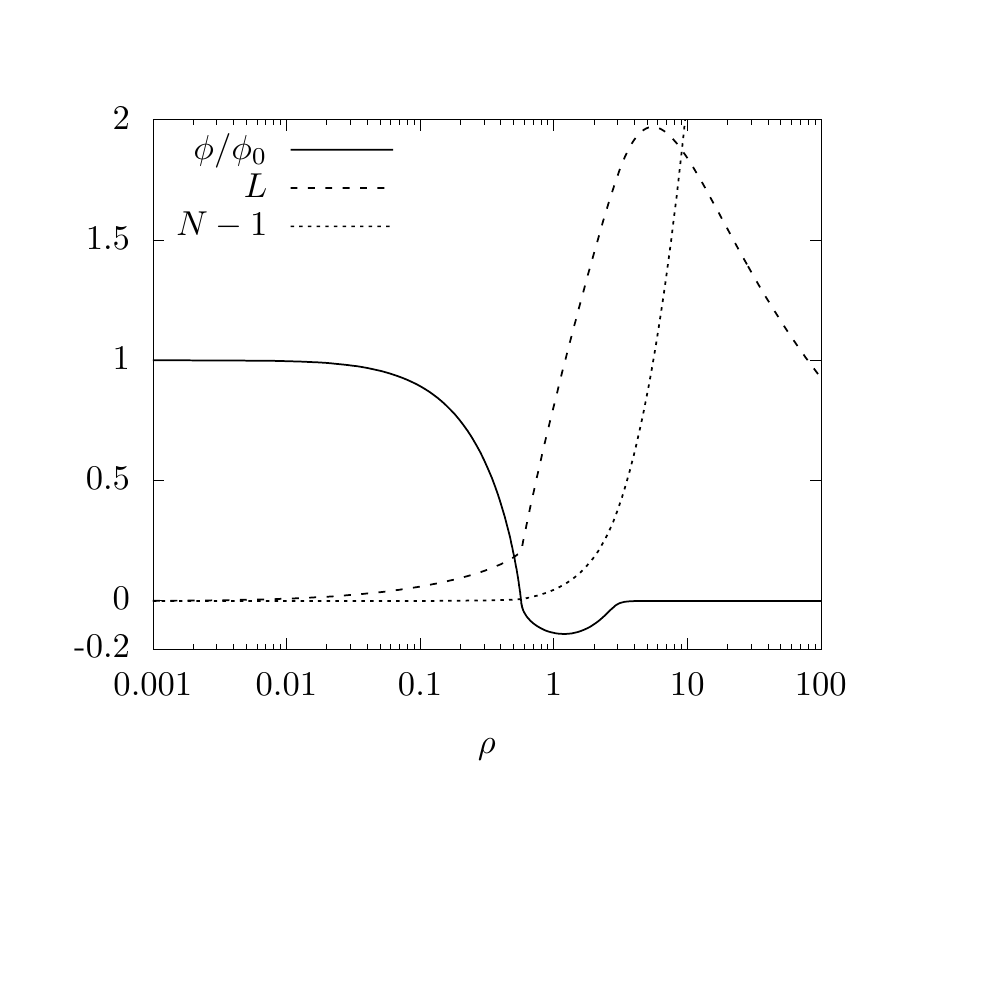}
\vspace{-2.5cm}
\begin{center}
\caption{We show the profiles of the scalar field function $\phi/\phi_0$ and the metric functions
$N$ and $L$ for a value of $\gamma$ close $\gamma_{\rm cr}$  for $B_0=1$ and $k=0$ (left) and $k=1$ (right), respectively.
\label{fig:gb_limiting}}
\end{center}
\end{figure}

Finally, we would like to discuss why the solutions chease to exist at $\gamma=\gamma_{\rm cr}$. 
A solution close to the limiting solution
is shown in Fig. \ref{fig:gb_limiting} for $k=0$ and $k=1$, respectively. Clearly, the scalar field becomes zero outside of a sharply defined
radius $\rho_{\rm cr}$ such that at $\rho=\rho_{\rm cr}$ the scalar field function is non-differentiable.
For $\rho > \rho_{\rm cr}$ the solution corresponds to the Melvin magnetic universe, while it possesses a non-trivial, scalarized interior.

\begin{figure}[h!]
\begin{center}
{\includegraphics[width=10cm]{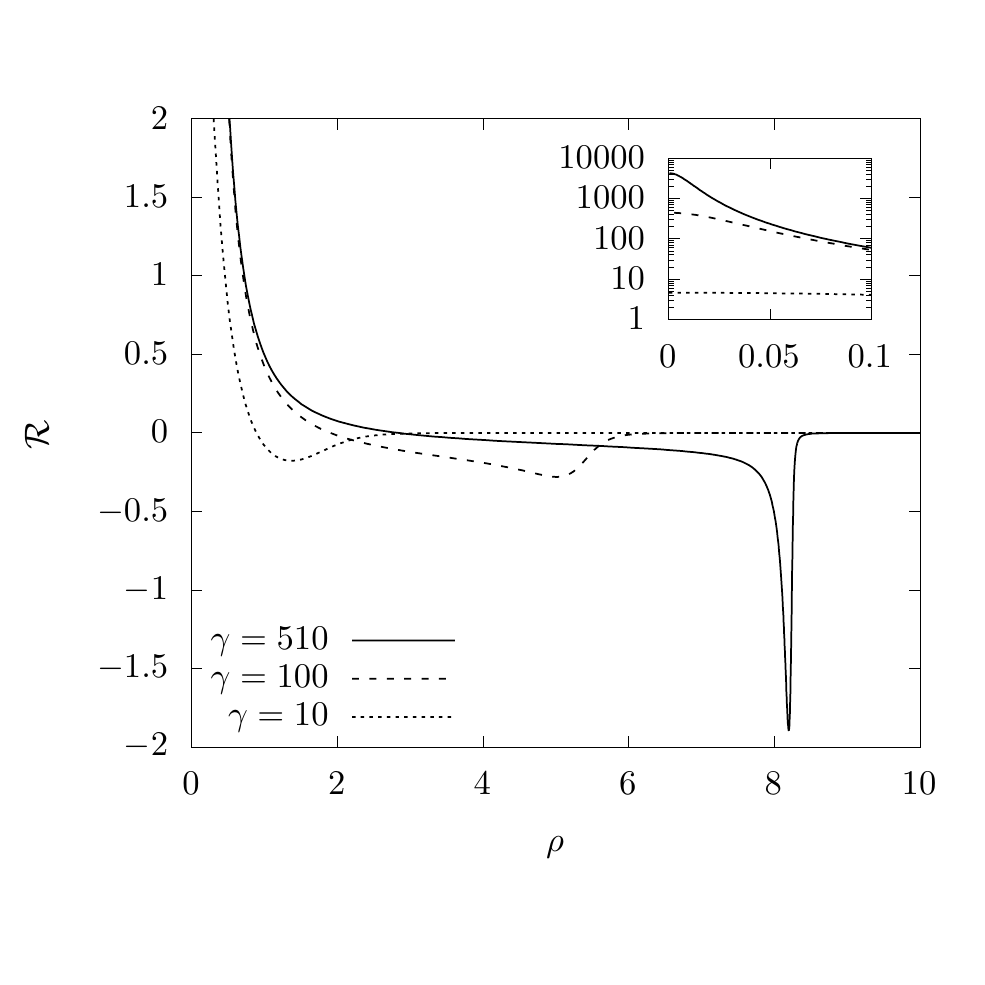}}
\vspace{-1cm}
\caption{We show the Ricci scalar ${\cal R}$ of the scalarized Melvin solution for $k=0$, $B_0=1$ and for different
values of $\gamma$ including $\gamma=510$ close to $\gamma_{\rm cr}$
\label{fig:ricci_limiting}
}
\end{center}
\end{figure}

We observe that the Ricci scalar ${\cal R}$ increases strongly on the symmetry axis of the solution when
approaching $\gamma_{\rm cr}$. This is demonstrated for the $k=0$ solution with $B_0=1$ in 
Fig.\ref{fig:ricci_limiting}, where we give ${\cal R}$ for $\gamma=10$ (close to $\gamma_0$), an intermediate
 $\gamma=100$ as well as for $\gamma=510$ (close to $\gamma_{\rm cr}$). The subfigure of Fig.\ref{fig:ricci_limiting} shows the strong increase of ${\cal R}$ at $\rho=0$. 
The figure further shows that the Ricci scalar becomes discontinuous at $\rho=\rho_{\rm cr}$ indicating that the limiting space-times possesses singularities.

\section{Conclusions}

In this paper, we have demonstrated that the process of {\it spontaneous scalarization} is not 
specific to compact objects such as black holes, neutron stars or boson stars, but also exists for
extended self-gravitating solutions. We have used the Melvin magnetic universe solution that describes
magnetic fields orientated into the direction of the symmetry axis and possesses a cylindrically
symmetric, static space-time. 
In the small scalar field limit, we find that the linear Klein-Gordon equation of a massive, real scalar field
leads to solutions described by the Laguerre polynomials close to the axis of symmetry and
by Bessel functions asymptotically, respectively. These results suggest that a discrete family of scalar field solutions exists in the model. We have demonstrated this by explicit numerical construction of the
solutions including backreaction of the space-time and the magnetic field, respectively. 
The solutions are characterized by two parameters: the magnetic field parameter $B_0$
and the value of the scalar field on the symmetry axis, which is linked to the value of the non-minimal coupling. 
The scalarized solutions, which are non-trivial deformations of the underlying Melvin magentic universe solution, exist only in specific intervals of the non-minimal couplings. 
Since these intervals for different node numbers are in some cases
mutually exclusive, the question arises whether the higher node solutions can be interpreted as
radial excitations of the fundamental solutions with no nodes, as is often done in self-gravitating systems.

It would be interesting to understand whether such scalarization processes of extended magnetic fields
could be observed in the universe and if not, if observations could provide limits on the coupling parameters
or even exclude extended gravity models with non-minimal coupling terms. 
An interesting future investigation could be another very specific cylindrically symmetric system that
possesses magnetic fields: the cosmic string. While the outside of a cosmic string is characterized by
a massive gauge field and a massive scalar field (spontaneously broken phase), the inside of the string
core remains in the symmetric, i.e. false vacuum of the model in which the gauge symmetry is unbroken.
It is surely of interest to understand whether cosmic strings that are hypothetical relics of the primordial
universe could be scalarized spontaneously and, if so, how this would change the properties of these objects.

\vspace{0.5cm}

{\bf Acknowledgments} R.C. thanks CAPES for financial support under grant No. {\it 88887.371717/2019-00}.  B. H. would like to thank FAPESP for financial support under grant No. {\it 2019/01511-5} as well as the DFG Research Training Group 1620 {\it Models of Gravity} for financial support.

\pagebreak


\end{document}